\documentclass[acmsmall,screen]{acmart}

\usepackage{tikz}
\usepackage{amsmath}
\usepackage{enumitem} 
\usepackage{amsmath}        
\usepackage{algorithm}              
\usepackage{algpseudocode}          
\usepackage{graphicx}               
\usepackage{caption}               
\usepackage{subcaption}            
\usepackage{tikz}               
\usepackage{booktabs}          
\usepackage{xcolor}                
\usepackage{enumitem}             
\usepackage{makecell} 
\usepackage{multirow}              
\usepackage[most]{tcolorbox}

\usepackage{amsmath}
\usepackage{graphicx}
\usepackage{booktabs}
\definecolor{eclipseGreen}{RGB}{63, 127, 95}  
\algrenewcommand{\algorithmiccomment}[1]{\hfill \textcolor{eclipseGreen}{\footnotesize // #1}}
\usepackage{filecontents}

\newcommand{\tool}{\textsc{Strike}}

\begin{document}

\title{Towards Effective Black-Box Adversarial Attacks on Deep Code Models via Structural and Identifier Perturbations}

\author{Bin Duan}
\affiliation{%
  \institution{The University of Queensland}
  \city{Brisbane}
  \country{Australia}}
\email{b.duan@uq.edu.au}

\author{Jintao Lin}
\affiliation{%
  \institution{The University of Queensland}
  \city{Brisbane}
  \country{Australia}}
\email{jintao.lin@student.uq.edu.au}

\author{Dan Dongseong Kim}
\affiliation{%
  \institution{The University of Queensland}
  \city{Brisbane}
  \country{Australia}}
\email{dan.kim@uq.edu.au}

\author{Guowei Yang}
\authornote{Corresponding author.}
\affiliation{%
  \institution{The University of Queensland}
  \city{Brisbane}
  \country{Australia}}
\email{guowei.yang@uq.edu.au}

\renewcommand{\shortauthors}{Duan et al.}

\begin{abstract}

Deep code models (DCMs) are increasingly embedded in code intelligence tasks. However, their robustness under adversarial attacks remains insufficiently understood. Prior black-box attacks mainly rely on identifier-only substitutions or structural edits transferred from reference samples, yielding perturbation spaces defined largely independently of the attacked input.
We introduce \tool, an input-conditioned black-box adversarial robustness-testing framework that constructs and searches a hierarchical perturbation space for each input.
\tool\ first partitions code into blocks and uses an LLM to generate context-specific structural candidates, filtered for syntactic validity, ranked by similarity, and adaptively combined.
It then refines the best structural variant through similarity-guided identifier substitutions drawn from dynamically constructed candidate pools.
Evaluations across representative code intelligence tasks, including clone detection, vulnerability detection, and code summarization, demonstrate that \tool\ achieves higher attack success rates with competitive target-model query overhead. Strategy-specific static checks across all three tasks and execution-based validation on the executable Juliet vulnerability-detection subset provide complementary, task-bounded evidence regarding the validity of the generated variants. Representation similarity and human evaluation further indicate that the variants remain highly similar to the original code and contextually natural. Fine-tuning with \tool-generated samples improves cross-attack performance on fixed perturbed evaluation sets while preserving clean performance.

\end{abstract}

\maketitle

\section{Introduction}

Recent progress in deep learning~(DL) has led to the emergence of deep code models~(DCMs)~\cite{le2020deep}, which are trained on large-scale source-code corpora~\cite{zhang2024review} and increasingly adopted for code intelligence tasks such as clone detection~\cite{ain2019systematic}, vulnerability identification~\cite{wang2020combining}, and code summarization~\cite{zhang2022survey}.
As in other DL domains~\cite{duan2025generating}, the robustness of these code-oriented models is essential~\cite{henkel2022semantic}. Prior studies~\cite{li2025importance, eleftheriadis2024adversarial} demonstrate that adversarial samples serve as valuable artifacts for examining and strengthening model robustness. By applying subtle perturbations to an input, one can produce adversarial samples that surface erroneous predictions~\cite{qu2024survey}. These adversarial samples can be used to expose erroneous predictions and further improve robustness through fine-tuning~\cite{zhao2022adversarial}. Therefore, generating adversarial samples is important for evaluating and improving the resilience of DCMs under perturbations.

However, transferring adversarial-generation strategies from other fields to code is non-trivial~\cite{qu2024survey}. 
Unlike continuous modalities such as images, source code is symbolic and governed by strict syntactic and structural constraints~\cite{yang2024srcmarker}. As a result, adversarial attacks are generally expected to remain syntactically valid and avoid disruptive changes~\cite{zhou2024evolutionary}. In practice, effective attacks also tend to be localized and contextually natural, so that the resulting samples remain plausible and useful for robustness evaluation~\cite{tipirneni2024structcoder}.

To this end, various methods have been proposed to generate adversarial samples for DCMs. Early black-box attacks, such as MHM~\cite{zhang2020mhm}, use heuristic search over identifier substitutions. Later identifier-oriented methods, including ALERT~\cite{yang2022alert}, BeamAttack~\cite{du2023beam}, and ITGen~\cite{huang2025itgen}, introduce more advanced candidate-generation and search strategies. ALERT employs genetic search to refine candidates, BeamAttack uses fixed-width beam search to prioritize key identifiers, and ITGen formulates the generation process as discrete Bayesian optimization to improve query efficiency. Despite their lexical plausibility, these approaches are restricted to identifier renaming, which may limit perturbation diversity and attack effectiveness. CODA~\cite{tian2023coda} combines structural and identifier edits transferred from opposite-label reference samples. Although this design extends the perturbation space beyond identifier renaming, its structural candidates are bounded by the reference pool, are not generated for the local context of the attacked block, and require class labels, which limits its applicability to generative tasks. These characteristics motivate a different formulation in which the perturbation space is constructed from the attacked input itself.
Building on these advances, we identify two key limitations:
\textbf{(i)} when limited to identifier substitutions, the adversarial search space becomes narrow and the generated samples tend to be homogeneous, limiting their ability to effectively evaluate model robustness; and
\textbf{(ii)} existing structural perturbations are either inflexible rule-based rewrites or context-misaligned reference transfers, both of which rely on predefined transformations whose perturbation space is determined independently of the input, often introducing inconsistencies with the original code and degrading contextual naturalness.
Taken together, existing methods construct their perturbation spaces largely independently of the attacked input, restricting contextual fit, task applicability, and the range of plausible variants that can be explored.

To address these limitations, we formulate black-box adversarial code generation as a hierarchical search over an input-conditioned perturbation space. Structural candidates are generated from the local context of each code block, while identifier candidates are constructed dynamically for each input; the two spaces probe complementary structural and lexical sensitivities. However, exhaustive search is costly and may yield unnatural or inconsistent candidates. 
We propose \tool, a coarse-to-fine black-box attack framework that performs sample-specific structural search via block partitioning and LLM-based rewrites, filters candidates based on syntactic validity and similarity, and adaptively combines them to identify high-impact perturbations. It then refines candidates via identifier substitutions with dynamically constructed pools to further reduce the task-specific attack score. This design produces syntactically valid perturbations with a broader range of variations, while tending to generate contextually plausible variants that are more useful for robustness evaluation.

We evaluate \tool\ using CodeBERT, CodeGPT, and CodeT5 on three representative tasks---clone detection, vulnerability detection, and code summarization---using standard benchmarks, and compare it against four state-of-the-art baselines: ALERT, BeamAttack, ITGen, and CODA. We report attack success rate, target-model query overhead, measured search time, and representation similarity, complemented by human evaluation.
Results show that \tool\ achieves higher attack success rates with competitive target-model query overhead. Strategy-specific static checks, representation similarity, and human evaluation provide complementary evidence regarding the validity and naturalness of the generated samples, while execution-based behavioral validation is additionally conducted on the executable Juliet subset. Adversarial fine-tuning with \tool-generated samples improves cross-attack performance on fixed perturbed evaluation sets while preserving clean performance. Ablation studies show that removing either stage degrades performance, and additional evaluations demonstrate compatibility with different LLM backends.
To sum up, this paper makes three contributions:

\noindent\textbf{Idea.}
We formulate black-box adversarial attacks on DCMs as hierarchical search over an input-conditioned perturbation space. Unlike identifier-only attacks and reference-transfer approaches, the perturbation space is constructed from the attacked program itself and does not require opposite-label samples, allowing the same formulation to support both classification and generation tasks.

\noindent\textbf{Technique.}
We instantiate this formulation in \tool, a coarse-to-fine framework that combines block-level LLM rewrites, syntax and similarity filtering, adaptive structural exploration, and dynamically pooled identifier substitutions. The two stages explore complementary structural and lexical perturbations while prioritizing syntactic validity and contextual fit.

\noindent\textbf{Evaluation.}
We conduct extensive experiments showing that \tool\ achieves higher attack success rates with competitive target-model query overhead. Strategy-specific static checks are applied across all three tasks, and execution-based behavioral validation is additionally conducted on the executable Juliet vulnerability-detection subset. Representation similarity and human evaluation provide complementary evidence regarding contextual naturalness and similarity. Adversarial fine-tuning results further show improved cross-attack performance on fixed perturbed evaluation sets while preserving clean performance.
To support reproducibility and future research, we released the implementation of \tool~\cite{Our}.

\section{Background}

Deep learning has been widely applied to source code processing~\cite{feng2020codebert,lu2021codegpt,wang2021codet5}. 
Several pre-trained models have been developed on large-scale code corpora, covering three mainstream architectures: encoder-only Transformer, like {CodeBERT}~\cite{feng2020codebert}, decoder-only Transformer, like {CodeGPT}~\cite{lu2021codegpt}, and encoder–decoder Transformer, like {CodeT5}~\cite{wang2021codet5}.
These models have driven substantial progress across a range of code-related tasks, including both {classification~\cite{combefis2022automated}} and {generation~\cite{chen2024survey}} tasks, by fine-tuning them on task-specific datasets. 
The former, such as clone detection~\cite{zhang2021survey} and vulnerability detection~\cite{fu2023chatgpt}, performs classification based on given code snippets~\cite{liu2022codebert}. While the latter, such as code summarization~\cite{zhang2024review} and code completion~\cite{izadi2024language}, generates descriptive or complementary sequences from code or natural language inputs.  
Following prior studies on adversarial sample generation for DCMs~\cite{du2023beam, huang2025itgen}, our work also focuses on these two representative tasks.

\begin{figure*}[t!]
  \centering
  \includegraphics[width=0.9\linewidth]{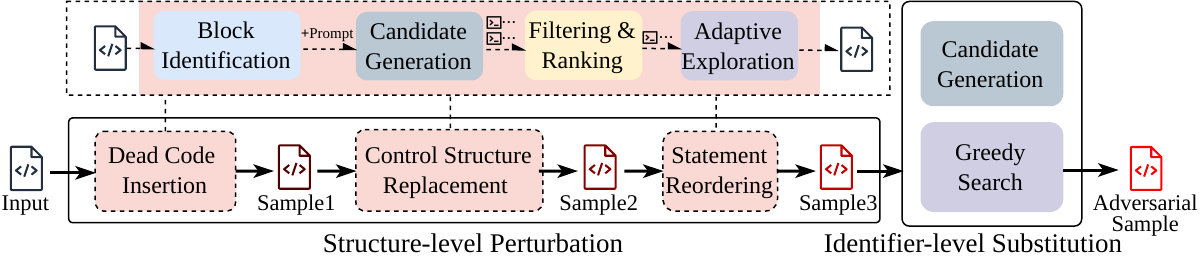}
  \vspace{-2mm}
  \caption{Overview of \tool.}
  \label{fig:overview}
  \vspace{-1mm}
\end{figure*}

\section{Threat Model}
\label{sec:threat_model}

\noindent\textbf{Attack surface and objective.}
We consider query-based black-box adversarial attacks on deep code models~(DCMs) in an offline benchmark setting. The target model is treated as externally deployed: the adversary can query its outputs but cannot access its parameters, gradients, training process, or private training data. Ground-truth labels for classification and reference summaries for generation are available to the benchmark evaluator for defining task-specific attack objectives. Accordingly, our primary goal is controlled robustness testing and improvement rather than demonstrating a fully label-free evasion attack against a real-world service.

\noindent\textbf{Adversary knowledge.}
We assume a score-based black-box setting for classification and an output-based black-box setting for generation, consistent with prior attacks on DCMs~\cite{yang2022alert,du2023beam,tian2023coda,huang2025itgen}. The adversary can query the target model but cannot access its parameters, gradients, or training process. For classification, the observable feedback consists of class confidence scores or probabilities. For generation, the observable feedback is the generated sequence, whose degradation is evaluated against the benchmark reference summary.
The adversary derives task-specific attack scores solely from these observable outputs.
The adversary may use publicly available auxiliary code corpora, static parsers, identifier embeddings, and off-the-shelf LLMs to construct candidate perturbations, consistent with prior black-box attacks on DCMs~\cite{yang2022alert,du2023beam,tian2023coda,huang2025itgen}.
These resources do not require access to the target model's parameters, gradients, training pipeline, or non-public data.

\noindent\textbf{Adversary capability.}
For clone detection, each input is a pair $[\mathrm{code}_1,\mathrm{code}_2,\mathrm{label}]$; for vulnerability detection, each sample is $[\mathrm{code}_1,\mathrm{label}]$; and for code summarization, each instance is $[\mathrm{code}_1,\mathrm{gold\_summary}]$.
Following prior adversarial attacks on DCMs, we set $x=\mathrm{code}_1$ as the attack target and keep $\mathrm{code}_2$, labels, and gold summaries unchanged.
The adversary may apply localized structure-level perturbations and identifier-level substitutions to $x$, but may not modify the target model, the training process, or any ground-truth annotation.

\noindent\textbf{Perturbation constraints.}
A candidate $x'$ is considered valid only if it satisfies the following constraints:
(i) it can be parsed into a syntactically valid program fragment;
(ii) identifier substitutions preserve lexical scope and binding consistency;
(iii) perturbations are localized and do not intentionally change the task-relevant behavior or annotation of the original sample; and
(iv) the resulting code remains contextually natural and close to the original code.
Because most benchmark datasets consist of isolated code snippets rather than complete executable projects, we do not claim formal semantic equivalence. Strategy-specific static checks are applied across all three tasks, while execution-based behavioral validation is conducted only on the executable Juliet vulnerability-detection subset. For clone detection and code summarization, the validity evidence is limited to static conformance checks, representation similarity, and human judgments, which provide complementary evidence but do not establish behavioral equivalence.

\noindent\textbf{Success criteria.}
Let $\mathcal{M}$ be the target DCM.
For classification tasks, we attack only samples that are originally correctly classified. 
Given the ground-truth label $y$, an attack succeeds if
\[
\arg\max_{y'} \mathcal{M}_{y'}(x') \neq y .
\]
During search, we minimize the confidence score assigned to the true label.
For code summarization, let $s$ denote the gold summary.
Following prior work~\cite{du2023beam,tian2023coda,huang2025itgen,yang2022alert}, an attack satisfies the established summarization criterion when its reference-overlap score reaches the predefined threshold:
\[
\mathrm{BLEU}\text{-}4(\mathcal{M}(x'), s) \leq \epsilon ,
\]
where we set $\epsilon=0$ in the main experiments, following prior work~\cite{du2023beam,huang2025itgen}.
This criterion measures the loss of four-gram overlap with the reference summary and enables direct comparison with prior attacks.

To unify the optimization process across classification and generation tasks, we define a task-specific attack score:
\[
A(x)=
\begin{cases}
p_{\mathcal{M}}(y \mid x), & \text{for classification tasks},\\
\mathrm{BLEU}\text{-}4(\mathcal{M}(x), s), & \text{for code summarization}.
\end{cases}
\]
A lower $A(x)$ indicates a stronger adversarial effect. Therefore, both structural exploration and identifier substitution are formulated as minimizing $A(x)$ under the perturbation constraints defined above.
For generation tasks, we evaluate attacks only on samples whose original BLEU-4 score is above the success threshold, i.e., 
\(\mathrm{BLEU}\text{-}4(\mathcal{M}(x),s)>\epsilon\). 
This avoids counting samples that already produce zero-overlap summaries before perturbation as successful attacks.

\section{Method}
\label{sec:method}

As shown in Fig.~\ref{fig:overview}, \tool\ operates in two stages: structure-level perturbation and identifier-level substitution. 
First, \tool\ applies a unified Identification--Generation--Filtering--Exploration pipeline to three LLM-based structural perturbation strategies: Dead Code Insertion, Control Structure Replacement, and Statement Reordering~(Section~\ref{sec:structure}). 
Starting from the original input, Dead Code Insertion identifies statement and control-structure blocks, generates candidates, filters and ranks them, and adaptively explores their combinations to obtain Sample~1.
Control Structure Replacement then takes Sample~1 as input and perturbs control-structure blocks to produce Sample~2, followed by Statement Reordering, which perturbs statement blocks of Sample~2 to produce Sample~3. 
This ordering expands the search space from low-risk insertions to more complex structural transformations.
If a successful adversarial sample is found at any stage, the process terminates early; otherwise, the best variant is passed to the next stage, where \tool\ dynamically constructs identifier substitution candidates and performs greedy renaming to further reduce the task-specific attack score and produce the final adversarial sample~(Section~\ref{sec:identifier}).

\subsection{LLM-based Structural Perturbations}
\label{sec:structure}

LLMs excel at producing developer-like code edits~\cite{mahmud2025enhancing}, which we leverage to introduce three structural perturbations.

\subsubsection{Dead Code Insertion}
\label{sec:DCI}
This perturbation strategy introduces diverse yet controlled structural variations by inserting developer-like statements generated by LLMs. 
It serves as the first structural perturbation strategy in our framework, consisting of four steps: code block identification, candidate generation, filtering and ranking, and adaptive exploration.

\begin{table*}[t]
\centering
\footnotesize
\caption{Prompts Used for LLM-based Structure-level Perturbations.}
\vspace{-2mm}
\label{tab:struct_prompts}
\setlength{\tabcolsep}{8pt}
\renewcommand{\arraystretch}{0.45}
\begin{tabular}{p{1.8cm} | p{11cm}}
\toprule
\textbf{Strategies} & \textbf{Prompt} \\
\midrule

\makecell[tl]{Dead Code\\Insertion} &
\parbox[t]{\linewidth}{\ttfamily
Insert 1--4 semantically inert local lines into the given code block.
The inserted lines should resemble developer-like temporary checks or local variable inspections, but must not perform I/O, call external methods, modify existing variables, change control flow, throw exceptions, or alter return values.
Reuse existing variables only in read-only expressions whenever possible.
Preserve the original code exactly when inserting, add only new lines, and return the full modified code without explanations or comments.
} \\
\midrule

\makecell[tl]{Control Structure\\Replacement} &
\parbox[t]{\linewidth}{\ttfamily
Rewrite this control structure into a semantically equivalent alternative form while preserving the original branch condition, loop bounds, body statements, side effects, and exception behavior.
Return only the complete modified code without explanations or comments.
} \\
\midrule

\makecell[tl]{Statement\\Reordering} &
\parbox[t]{\linewidth}{\ttfamily
Reorder only mutually independent statements in the following code snippet. Preserve every original statement exactly. Do not add, remove, or modify any statement. Return only the complete reordered code, without explanations or comments.
} \\
\bottomrule
\end{tabular}
\vspace{-1mm}
\end{table*}

\noindent \textbf{Code Block Identification.}
We begin by statically parsing the input code to detect syntactic boundaries and segment it into fine-grained {code blocks}, which serve as the fundamental units for structure-level perturbations.
The parsing process is implemented using {Tree-sitter}~\cite{brunsfeld2018tree-sitter}, a parsing framework that supports multiple programming languages through language-specific parsers.
Each block is treated as an independent region where perturbations can be applied locally, enabling the LLM to generate precise modifications within well-defined boundaries while reducing unintended changes to the original code.

We identify two types of code blocks using Tree-sitter: {Statement blocks} and {Control-structure blocks}. 
{Statement blocks} include declarations and expression statements. 
In Java-like languages: \texttt{local variable declarations}, \texttt{assignment expressions}, and \texttt{method invocations}. 
In C-like languages: \texttt{declarations}, \texttt{assignment expressions}, and \texttt{call expressions}. 
{Control-structure blocks} include conditionals and loops. 
In Java-like languages: \texttt{if}, \texttt{switch}, \texttt{for}, \texttt{while}, \texttt{do}, \texttt{enhanced for}, and \texttt{try}; 
In C-like languages: \texttt{if}, \texttt{switch}, \texttt{for}, \texttt{while}, and \texttt{do}.

Only the lowest-level nested blocks are retained to avoid overlapping and redundant hierarchical perturbations.
For Dead Code Insertion, we identify both blocks, as this perturbation aims to introduce lightweight, contextually neutral insertions throughout the code without altering its logical structure.
Including both block types ensures broad coverage of potential insertion points, allowing LLMs to generate developer-like edits.

\noindent \textbf{Candidate Generation.}  
Dead Code Insertion generates candidate variants by injecting a small number of lines into each identified block while preserving syntactic validity and avoiding disruptive changes. 
The objective is to produce natural, developer-like candidates that increase local complexity without disrupting the overall structure, thereby testing the model’s sensitivity to structural variations.
For each code block \(b \in \mathcal{B}\), we query the LLM using a strategy-specific prompt, as shown in Table~\ref{tab:struct_prompts}, of the form
\(
\text{\texttt{[prompt] + [code block $b$]}}.
\)
The prompt instructs the model to insert 1--4 lines resembling temporary debugging or inspection statements, prohibits any modification or deletion of original lines, and requires returning the complete modified block without explanations.  
For each block, one LLM request generates \(n\) candidate rewrites, denoted as \(\mathcal{G}(b)=\{g_1,g_2,\ldots,g_n\}\), where \(g_i\) is the \(i\)-th candidate for block \(b\).
All outputs of LLMs are designed as drop-in replacements, aiming to maintain structural similarity and contextual naturalness of the original code, while ensuring syntactic correctness through static validation.
The union of all per-block candidate sets forms the global candidate space: \(\mathcal{N} = \bigcup_{b \in \mathcal{B}} \mathcal{G}(b),\)
which serves as the initial input for the subsequent filtering and exploration steps.

\noindent \textbf{Filtering and Ranking.}  
All generated candidates are first deduplicated and statically validated; candidates that fail parsing are discarded.
Static validation is performed by detecting AST parsing errors using Tree-sitter~\cite{brunsfeld2018tree-sitter}, retaining only syntactically correct candidates.
For each code block \(b \in \mathcal{B}\), \(\mathcal{G}(b) = \{g_1, g_2, \ldots, g_n\}\) denote the set of candidates generated by LLMs.  
We compute embedding similarity between each candidate and the original block to prioritize candidates that remain structurally consistent with the input:
\[
s_i = \cos\bigl(\mathrm{Emb}(b),\, \mathrm{Emb}(g_i)\bigr), \quad g_i \in \mathcal{G}(b),
\]
where \(\mathrm{Emb}(\cdot)\) denotes the code embedding function, obtained by mean pooling over token representations from a pretrained masked language model~\cite{feng2020codebert}, and \(s_i \in [-1,1]\) measures embedding similarity.
Candidates are ranked in descending order of \(s_i\), and the top-\(k\) candidates are retained for further exploration:
\[
\mathcal{G}'(b) = \text{Top-}k\bigl(\mathcal{G}(b),\, s_i\bigr).
\]
The union of these per-block refined sets forms the global candidate pool:
\(
\mathcal{N}' = \bigcup_{b \in \mathcal{B}} \mathcal{G}'(b),
\)
which filters out invalid candidates, prioritizes structurally consistent variants, and reduces the search space for the next step. Additionally, each candidate preserves its original position and indentation, ensuring precise alignment and improving code naturalness.

\begin{algorithm}[t!]
\caption{Adaptive Exploration}
\footnotesize
\label{alg:adaptive_search}
\begin{algorithmic}[1]
\Require Refined candidate pool $\mathcal{N}'$, target model $\mathcal{M}$, task-specific attack score $A(\cdot)$
\State Initialize $\mathcal{C} = \{c_1,\ldots,c_{r_0}\}$ from $\mathcal{N}'$
\State $r = r_0$; $A_{\text{orig}} = A(\text{input code})$; $A^\star = A_{\text{orig}}$; $p = 0$; $c^\star = \text{input code}$
\While{$\mathcal{C}\neq\emptyset$}
  \ForAll{$c\in\mathcal{C}$}
    \State Evaluate $A(c)$ using $\mathcal{M}$; $\Delta(c) = A^\star - A(c)$
    \If{$c$ satisfies the attack success criterion}
      \State \Return $c$
    \EndIf
  \EndFor
  \State $c^* = \arg\max_{c\in\mathcal{C}}\Delta(c)$
  \If{$A(c^*) < A^\star$}
    \State $A^\star = A(c^*)$; $c^\star = c^*$; $p = 0$
    \State Replace $\mathcal{C}$ with up to $r_0$ newly generated candidates obtained by applying previously unexplored block-level perturbations to $c^*$
    \State $r = r_0$
\Else
  \State $p = p + 1$
  \If{$p \ge q$}
    \If{$r \le 1$}
      \State \Return $c^\star$
    \EndIf
    \State $r = r - 1$; $p = 0$
  \EndIf
  \State Replace $\mathcal{C}$ with up to $r$ newly generated candidates obtained by applying previously unexplored block-level perturbations to $c^\star$
\EndIf
\EndWhile
\State \Return $c^\star$
\end{algorithmic}
\end{algorithm}

\noindent \textbf{Adaptive Exploration.}
Algorithm~\ref{alg:adaptive_search} performs adaptive exploration to identify the most adversarially effective complete-code candidate. 
It first initializes the active candidate set $\mathcal{C}$ from the refined pool $\mathcal{N}'$, and sets the initial search size $r=r_0$, the baseline attack score $A_{\text{orig}}$, the current best score $A^\star$, and the patience counter $p$.
The algorithm then iteratively evaluates all active candidates using the task-specific attack score $A(\cdot)$ and selects the candidate with the largest score reduction as the most promising direction.
If a candidate already satisfies the task-specific attack success criterion, the search terminates early.
Otherwise, if the best candidate improves over the current best score, the algorithm updates the best sample and replaces the active candidate set with newly generated neighbors obtained from previously unexplored block-level perturbations. Thus, the candidate set changes after each successful expansion, and previously explored block-level candidates are not repeatedly applied to the same search state.
When an evaluated batch yields no improvement, the algorithm refreshes the active set using previously unexplored perturbations of the best sample found so far. The patience counter therefore records consecutive unsuccessful candidate batches rather than repeated evaluations of the same candidates. After $q$ unsuccessful batches, the active search width $r$ is reduced by one; the search terminates when no further narrowing is possible.

\subsubsection{Control Structure Replacement}
\label{sec:CSR}
The second structural perturbation strategy rewrites control structures into alternative forms to introduce syntactic variations.
Starting from Sample~1, we identify only control-structure blocks, excluding statement blocks. 
For each block, the LLM is prompted to modify only the control handle while preserving the block body (Table~\ref{tab:struct_prompts}). 
The generated candidates are then filtered, ranked, and explored using the same procedure as in Section~\ref{sec:DCI}. 
The best resulting variant is recorded as Sample~2 in Fig.~\ref{fig:overview}.

\subsubsection{Statement Reordering}
\label{sec:SR}
The third structural perturbation strategy reorders independent statements into alternative forms.
Starting from Sample~2, we identify only local statement blocks, excluding control structures, so that reordering is restricted to statements that are independent and mainly affects surface-level execution order. 
To control block length, statement sequences longer than ten lines are split into smaller blocks. 
For each block, the LLM is prompted to reorder only independent statements. 
The generated candidates are then filtered, ranked, and explored following the same procedure as in the previous steps, yielding Sample~3 in Fig.~\ref{fig:overview}.

To reduce the risk of changing program behavior, we additionally apply a conservative dependency filter. 
For each reordered candidate, we require that the multiset of original statements is unchanged and that statements with apparent def-use dependencies preserve their original relative order. 
Candidates violating these constraints or failing Tree-sitter parsing are discarded.

\subsection{Similarity-Guided Identifier Substitution}
\label{sec:identifier}

Building upon structure-level perturbation, we further perform identifier-level substitutions to reduce the task-specific attack score while preserving lexical scope and prioritizing lexically similar replacements.
Following previous works~\cite{yang2022alert,du2023beam,huang2025itgen}, this step focuses on systematically renaming variables and functions, while ensuring consistent bindings across all occurrences.  
Because identifier renaming typically does not alter the underlying logic or data flow when consistency is maintained, such perturbations introduce lexical variations that expose the model’s sensitivity to superficial naming patterns.
We implement this process through two steps: Candidate Generation and Iterative Greedy Search. \tool\ constructs a similarity-guided candidate pool for each identifier. We first retrieve the {least similar} code snippets from the public auxiliary corpus for each input, and collect all identifiers appearing in them as a global candidate pool. 
Then, for each renamable identifier, we rank candidate substitutions by lexical similarity and retain those with higher similarity, promoting more plausible naming patterns while enabling effective adversarial perturbations.

\noindent \textbf{Candidate Generation.}  
We begin by extracting the complete set of identifiers $\mathcal{I} = \{i_1, i_2, \ldots, i_m\}$ from the input code using the Tree-sitter~\cite{brunsfeld2018tree-sitter} parser to build an accurate AST and record each identifier’s syntactic role and lexical scope.  
This ensures that only renameable identifiers are selected and that substitutions remain consistent across references without introducing inconsistencies.
Next, to construct the substitution candidate set, we first sample $\tau$ external code from the public auxiliary corpus and denote the sampled set by
\(
\mathcal{D}=\{d_1,d_2,\ldots,d_\tau\}.
\)
We then compute the code-level similarity between each $d\in\mathcal{D}$ and the input code using cosine averaging on hidden-layer embeddings based~\cite{feng2020codebert}:
\[
S(d) = \frac{1}{L} \sum_{t=1}^{L} \cos\bigl(h_t(\text{input}),\, h_t(d)\bigr),
\]
where $h_t(\cdot)$ is the final hidden representation at token position $t$, and $L$ is the aligned sequence length~(up to 512, with truncation or padding applied as needed).  
To enhance the adversarial strength of identifier substitutions, we first retain the $\phi$ least similar code:
\[
\mathcal{D}' = \text{Bottom-}\phi\bigl(\mathcal{D},\, S(d)\bigr).
\]
For each selected external program $d \in \mathcal{D}'$, we extract its identifier set 
\(\mathcal{E}(d) = \{e_1, e_2, \ldots, e_{n}\}\), 
which contains all identifiers appearing in $d$.  
All such sets are then aggregated to construct the global external identifier pool:
\(
\mathcal{E} = \bigcup_{d \in \mathcal{D}'} \mathcal{E}(d),
\)
which serves as the substitution set.
For each renamable identifier $i \in \mathcal{I}$, we compute lexical similarity with every external identifier $e \in \mathcal{E}$ using FastText~\cite{bojanowski2017enriching} embeddings:
\[
s(e \mid i) = \cos\bigl(\mathbf{v}(i),\, \mathbf{v}(e)\bigr),
\]
where $\mathbf{v}(\cdot)$ denotes the FastText vector.
We then retain the top-$\eta$ most similar external identifiers as the substitution candidate set for $i$:
\[
\mathcal{R}(i) = \text{Top-}\eta\bigl(\mathcal{E},\, s(e \mid i)\bigr).
\]
Thus, we construct a per-identifier substitution mapping
\(
\mathcal{R} = \{\mathcal{R}(i) \mid i \in \mathcal{I}\},
\)
where each identifier $i$ is associated with a set of $\eta$ lexically similar candidates. Candidates are drawn from a dynamically constructed identifier pool to enhance adversarial effectiveness, while lexical similarity helps preserve naming naturalness and readability.

\noindent \textbf{Iterative Greedy Search.}  
We then apply identifier substitutions to further reduce the task-specific attack score.
At each step, the algorithm evaluates candidate renamings and greedily selects the substitution that yields the largest score reduction, ensuring consistent bindings across all identifier occurrences.
This process proceeds until no additional performance degradation is achieved or early-stopping criteria are triggered, after which the final Adversarial Sample is produced (Fig.~\ref{fig:overview}).

\begin{algorithm}[t]
\caption{Greedy Identifier Substitution}
\label{alg:identifier_greedy}
\footnotesize
\begin{algorithmic}[1]
\Require Identifier set $\mathcal{I}$, substitution candidates $\mathcal{R}$, target model $\mathcal{M}$, input code, task-specific attack score $A(\cdot)$
\State $\text{adv\_code}=\text{input}$; $A_{\mathrm{cur}}=A(\text{input})$
\State $\text{best\_code}=\text{adv\_code}$; $A_{\mathrm{best}}=A_{\mathrm{cur}}$
\State Mark all $i\in\mathcal{I}$ as unconfirmed; $t=0$

\While{$\exists\, i\in\mathcal{I}$ unconfirmed}
  \State $A_{\mathrm{cand}}=+\infty$; $(i^\star,r^\star)=(\emptyset,\emptyset)$

  \For{each unconfirmed $i\in\mathcal{I}$}
    \For{each $r\in\mathcal{R}(i)$}
      \State $\text{tmp\_code}=\text{adv\_code with }i\rightarrow r$
      \State Evaluate $A(\text{tmp\_code})$ using $\mathcal{M}$

      \If{$\text{tmp\_code}$ satisfies the attack success criterion}
        \State \Return $\text{tmp\_code}$
      \EndIf

      \If{$A(\text{tmp\_code})<A_{\mathrm{cand}}$}
        \State $A_{\mathrm{cand}}=A(\text{tmp\_code})$
        \State $(i^\star,r^\star)=(i,r)$
      \EndIf
    \EndFor
  \EndFor

  \If{$i^\star=\emptyset$}
    \State \Return $\text{best\_code}$
  \EndIf

  \State $\Delta=A_{\mathrm{cur}}-A_{\mathrm{cand}}$
  \State $\text{adv\_code}=\text{adv\_code with }i^\star\rightarrow r^\star$
  \State $A_{\mathrm{cur}}=A_{\mathrm{cand}}$; mark $i^\star$ confirmed

  \If{$A_{\mathrm{cur}}<A_{\mathrm{best}}$}
    \State $A_{\mathrm{best}}=A_{\mathrm{cur}}$; $\text{best\_code}=\text{adv\_code}$
  \EndIf

  \If{$\Delta>0$}
    \State $t=0$
  \Else
    \State $t=t+1$
    \If{$t\ge\kappa$}
      \State \Return $\text{best\_code}$
    \EndIf
  \EndIf
\EndWhile

\State \Return $\text{best\_code}$
\end{algorithmic}
\end{algorithm}

Algorithm~\ref{alg:identifier_greedy} performs identifier substitution through a greedy search with limited non-improving moves. At each iteration, the algorithm evaluates all candidate substitutions for the remaining unconfirmed identifiers and selects the candidate with the lowest task-specific attack score. If the selected substitution improves the current score, the patience counter is reset. Otherwise, the substitution is still committed to change the search state and explore identifier combinations beyond the current local optimum, and the patience counter is incremented. The search terminates after $\kappa$ consecutive non-improving substitutions. The algorithm separately maintains the current search state and the best variant observed so far, and returns the latter if no successful adversarial sample is found.

\section{Evaluation}

\subsection{Research Questions}
Our evaluation aims to answer the following research questions:
\begin{itemize}
  \item[\textbf{RQ1:}] How effective and efficient is \tool\ in attacking deep code models across different tasks?
  \item[\textbf{RQ2:}] To what extent do the adversarial samples generated by \tool\ preserve task-relevant behavior while maintaining high similarity and contextual naturalness?
  \item[\textbf{RQ3:}] How do adversarial samples affect model robustness when used for adversarial fine-tuning?
  \item[\textbf{RQ4:}] How do the individual components of \tool\ contribute to its overall effectiveness?
\end{itemize}

For \textbf{RQ1}, we comprehensively evaluate the effectiveness and efficiency of \tool\ in generating adversarial samples across multiple tasks and models, and compare it with state-of-the-art baselines.  
For \textbf{RQ2}, we assess the extent to which the generated adversarial samples preserve task-relevant behavior while remaining similar to the original code and contextually natural. We apply strategy-specific static checks across all three tasks, execution-based behavioral validation on the executable Juliet subset, and complementary representation-similarity and human evaluations.
For \textbf{RQ3}, we further examine the practical value of generated adversarial samples by using them for adversarial fine-tuning, and analyze whether such training improves robustness to perturbed inputs. 
For \textbf{RQ4}, we perform ablation studies to isolate the impact of each key component of \tool. 
We evaluate different LLMs for structure-level perturbations and remove individual modules of the framework to quantify their contributions.

\subsection{Experimental Setup}

\begin{table}[t!]
\centering
\footnotesize
\caption{Statistics of Our Used Subjects.}
\vspace{-2mm}
\label{tab:subjects}
\renewcommand{\arraystretch}{0.85}
\begin{tabular}{
    p{3cm}  
    p{3cm}  
    p{3cm}  
    p{3cm}  
}
\toprule
\textbf{Task} & \textbf{Train/Val/Test} & \textbf{Model} & \textbf{Acc.(\%)/BLEU-4} \\
\midrule
\makecell[l]{Clone \\ Detection} 
  & 90,102/4,000/4,000 
  & \makecell[l]{CodeBERT\\CodeGPT\\CodeT5}
  & \makecell[c]{96.32\%\\96.52\%\\96.85\%} \\
\midrule
\makecell[l]{Vulnerability \\ Detection} 
  & 21,854/2,732/2,732 
  & \makecell[l]{CodeBERT\\CodeGPT\\CodeT5}
  & \makecell[c]{63.76\%\\65.23\%\\63.62\%} \\
\midrule
\makecell[l]{Code \\ Summarization} 
  & 164,923/5,183/10,955 
  & \makecell[l]{CodeBERT\\CodeGPT\\CodeT5}
  & \makecell[c]{18.75\\19.68\\20.40} \\
\bottomrule
\end{tabular}
\footnotesize{* BLEU-4 is reported for summarization; Acc.(\%) for the other tasks.}
\end{table}

\begin{table*}[t!]
\centering
\footnotesize
\caption{ASR of Different Attack Methods on Three Tasks.}
\vspace{-2mm}
\label{tab:asr}
\setlength{\tabcolsep}{2pt}
\renewcommand{\arraystretch}{0.5}
\begin{tabular}{l ccccc|ccc|ccc}
\toprule
\multirow{2}{*}{\textbf{Model}} 
& \multicolumn{5}{c}{\textbf{Clone Detection}} 
& \multicolumn{3}{c}{\textbf{Vulnerability Detection}}
& \multicolumn{3}{c}{\textbf{Code Summarization}} \\
\cmidrule(lr){2-6} \cmidrule(lr){7-9} \cmidrule(lr){10-12}
& \tool & CODA & ITGen & BeamAttack & ALERT
& \tool & CODA  & ALERT
& \tool & BeamAttack & ALERT \\
\midrule
\textbf{CodeBERT} 
& \textbf{41.98\%} & 32.37\% & 27.13\% & 21.60\% & 22.82\%
& \textbf{98.88\%} & 89.12\% & 50.49\% 
& \textbf{89.98\%} & 48.74\% & 62.50\%  \\
\textbf{CodeGPT} 
& \textbf{32.45\%} & 18.24\% & 15.47\% & 9.62\% & 12.49\%
& \textbf{98.82\%} & 94.95\% & 62.46\% 
& \textbf{66.31\%} & 29.37\% & 39.91\%  \\
\textbf{CodeT5} 
& \textbf{32.07\%} & 23.05\% & 19.36\% & 11.27\% & 18.24\%
& \textbf{97.91\%} & 95.88\% & 76.36\%
& \textbf{64.25\%} & 33.66\% & 42.93\%  \\
\midrule
\textbf{Average}
& \textbf{35.50\%} & 24.55\% & 20.65\% & 14.16\% & 17.85\%
& \textbf{98.54\%} & 93.32\% & 61.10\%
& \textbf{73.51\%} & 37.26\% & 48.44\% \\
\bottomrule
\end{tabular}
\vspace{-2mm}
\end{table*}

\noindent\textbf{Tasks and datasets.}
Table~\ref{tab:subjects} summarizes the tasks, datasets, and models. 
Following prior works~\cite{du2023beam,huang2025itgen}, we assess \tool\ on three representative deep code tasks covering both classification and generation: clone detection, vulnerability detection, and code summarization.
These tasks span both C and Java, enabling evaluation across multiple languages and tasks.
For clone detection, we use BigCloneBench~\cite{svajlenko2014bigclonebench}, which provides labeled pairs of functionally similar code snippets.
For vulnerability detection, we use the Juliet C suite~\cite{boland2012juliet}, where each function is annotated as vulnerable or benign.
For code summarization, we use CodeSearchNet~\cite{husain2019codesearchnet}, where the task is to generate a natural-language summary.

\noindent\textbf{Models.}
We evaluate three representative DCMs widely used in prior work~\cite{yang2022alert, du2023beam, tian2023coda}: CodeBERT~\cite{liu2022codebert} (encoder-only), CodeGPT~\cite{lu2021codegpt} (decoder-only), and CodeT5~\cite{wang2021codet5} (encoder--decoder), covering diverse architectures and serving as standard benchmarks with performance aligned to prior work.
We report Accuracy(\%) for classification and BLEU-4 for generation.

\noindent\textbf{Baselines.} 
We compare \tool\ with four representative state-of-the-art black-box attacks on DCMs: ALERT~\cite{yang2022alert}, BeamAttack~\cite{du2023beam}, CODA~\cite{tian2023coda}, and ITGen~\cite{huang2025itgen}. 
These methods cover different paradigms, including identifier-based attacks (ALERT, BeamAttack, ITGen) and hybrid structure–identifier attacks (CODA). 
Following common practice in adversarial code attacks, each baseline is evaluated on tasks where it is applicable. 
Specifically, ITGen provides an open-source implementation only for clone detection and is therefore evaluated on that task only. 
CODA constructs candidates from opposite-label samples and is thus not applicable to the unlabeled code summarization task. 
The released BeamAttack implementation relies on a preprocessed dataset with sectioned identifiers and does not provide the corresponding parsing and ordering pipeline, preventing its migration to the Juliet dataset.
All baselines are run using their official implementations and recommended default configurations, following their original experimental protocols~\cite{yang2022alert,du2023beam,tian2023coda,huang2025itgen}.

\noindent\textbf{Configuration.} 
For structure-level perturbations, each LLM request generates $n{=}10$ candidate rewrites for one block, of which the top $k{=}5$ are retained based on syntactic validity and structural similarity, with $r_0{=}3$ active candidates and patience threshold $q{=}2$.
For identifier-level substitutions, we sample $\tau{=}256$ external code, select the $\phi{=}32$ least similar ones, and retain $\eta{=}30$ lexically similar candidates for greedy search, with patience $\kappa{=}2$.
All experiments run on Ubuntu 24.04.3 LTS with a 64-core CPU and an NVIDIA RTX 6000 Ada GPU.

\noindent\textbf{Auxiliary Corpus.}
For identifier substitution, the auxiliary corpus is constructed from the publicly available non-test split of each benchmark, following prior black-box attacks that mine identifier candidates from public benchmark corpora~\cite{yang2022alert,tian2023coda}.
It is used only to obtain candidate identifier names.
Attack generation does not access the target model's parameters, gradients, architecture internals, training pipeline, or non-public data.

\noindent\textbf{LLMs.} 
We use both open-source and closed-source LLMs, OpenChat-3.5~\cite{wang2023openchat} and GPT-5-nano~\cite{openai_gpt5nano}, to verify compatibility across models. By default, OpenChat-3.5 is used for comparisons with baselines.

\noindent\textbf{Statistical Analysis.}
We compare \tool\ with each applicable baseline using two-sided paired permutation tests over per-input results averaged across the three runs.
Holm correction is applied to account for multiple comparisons, with statistical significance determined at $\alpha=0.05$.

\subsection{Metrics}
\label{metrics}

We evaluate the generated adversarial samples along three dimensions: effectiveness, efficiency, and similarity.

\noindent\textbf{\textit{Attack Success Rate (ASR)}}, widely used in~\cite{yang2022alert,du2023beam,tian2023coda,huang2025itgen}, measures the proportion of originally correctly handled samples for which an attack satisfies the task-specific success criterion.
For classification tasks, success means that the predicted label differs from the ground-truth label.
For code summarization, following prior work~\cite{yang2022alert,du2023beam,tian2023coda,huang2025itgen}, success means that BLEU-4 between the generated summary and the gold summary drops to 0.
For \tool, a final variant is counted as a successful attack only if it passes all applicable strategy-specific static checks.

\noindent\textbf{\textit{Average Model Queries (AMQ)}}, widely used in~\cite{yang2022alert,du2023beam,tian2023coda,huang2025itgen}, reports the average number of target-DCM invocations, computed as the total target-model queries divided by the number of attacked samples.

\noindent\textbf{\textit{Average Running Time (ART)}}, following prior work~\cite{yang2022alert,du2023beam,tian2023coda,huang2025itgen}, measures the average end-to-end wall-clock time required to attack one sample. It includes candidate generation, candidate parsing, filtering and ranking, target-model inference, and structural and identifier search.

\noindent\textbf{\textit{Average Code Similarity (ACS)}}, measures similarity between original and adversarial code in embedding space, computed as the cosine similarity of representations~\cite{feng2020codebert}. 



\section{Results and Analysis}
To mitigate randomness, each experiment was repeated three times and the average results are reported, except for the human study. Statistical comparisons follow the protocol described in the experimental setup.

\subsection{RQ1: Effectiveness and Efficiency}

\begin{table}[t]
\centering
\footnotesize
\caption{Class-Wise ASR on Vulnerability Detection.}
\vspace{-1mm}
\label{tab:classwise_asr}
\setlength{\tabcolsep}{4pt}
\renewcommand{\arraystretch}{0.8}
\begin{tabular}{lccccc}
\toprule
\textbf{Model} &
\textbf{Vuln. $n$} &
\textbf{V$\rightarrow$B} &
\textbf{Benign $n$} &
\textbf{B$\rightarrow$V} &
\textbf{Overall} \\
\midrule
CodeBERT & 905 & 99.23\% & 837 & 98.50\% & 98.88\% \\
CodeGPT  & 944 & 99.15\% & 838 & 98.45\% & 98.82\% \\
CodeT5   & 891 & 98.54\% & 847 & 97.25\% & 97.91\% \\
\bottomrule
\end{tabular}
\vspace{-2mm}
\end{table}

\subsubsection{Effectiveness}

For classification tasks, we report the proportion of label changes.
For code summarization, following prior work~\cite{du2023beam,huang2025itgen}, we report the proportion of cases where BLEU-4 with the reference summary drops to 0.
This criterion quantifies severe reference-overlap degradation and provides a common basis for comparing the evaluated attacks.
As shown in Table~\ref{tab:asr}, \tool\ achieves the highest ASR across all tasks and models. 
The improvement is most pronounced on clone detection, where \tool\ outperforms both transfer-based attacks (CODA) and identifier-only methods (ITGen, BeamAttack, ALERT). 
This advantage stems from the broader adversarial search space explored by \tool: identifier-based attacks are limited to lexical renaming, while transfer-based approaches rely on opposite-label samples. 
In contrast, \tool\ generates input-conditioned structural perturbations and further refines them with identifier substitutions. 
A similar trend appears in vulnerability detection and code summarization. 
The clean accuracy on Juliet ranges from 63.62\% to 65.23\%, and the standard ASR protocol considers only originally correctly classified inputs.
Table~\ref{tab:classwise_asr} therefore reports the eligible vulnerable and benign samples separately, together with the corresponding V$\rightarrow$B and B$\rightarrow$V attack success rates.
Identifier-only attacks perform particularly poorly on summarization, indicating that structural perturbations provide substantial additional attack effectiveness in the evaluated generation settings.
Overall, these results indicate that combining structural and identifier perturbations provides a more effective attack.
The improvements over the strongest applicable baseline are statistically significant in 9 of 9 task--model settings after Holm correction ($p<0.05$).

\begin{figure}[t!]
    \centering
    \includegraphics[width=\linewidth]{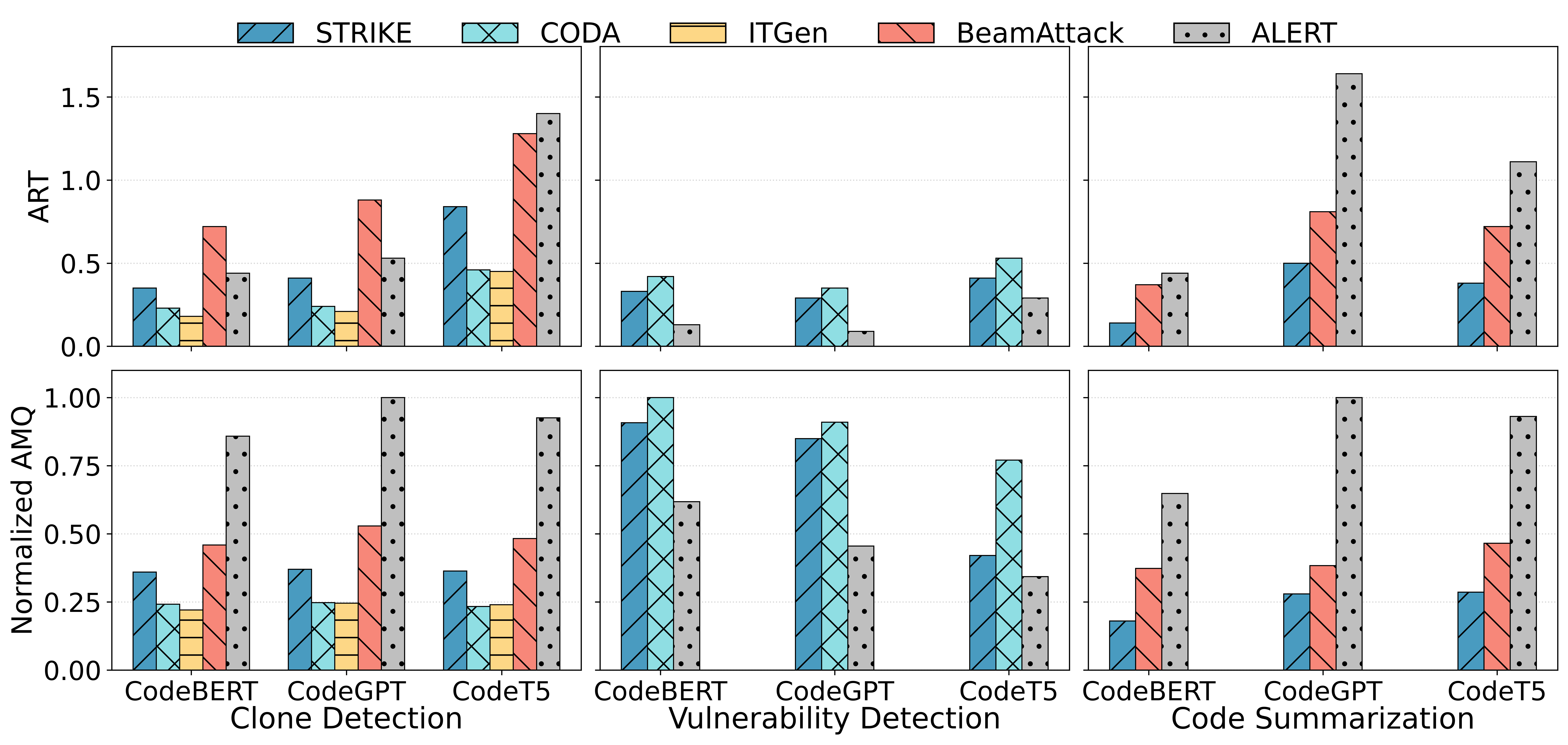}
    \vspace{-3mm}
    \caption{Comparison in Terms of ART and AMQ.}
    \label{fig:AMQ_ART}
    \vspace{-3mm}
\end{figure}

\begin{figure}[t!]
    \centering

    \begin{minipage}{\linewidth}
        \centering
        \includegraphics[width=\linewidth]{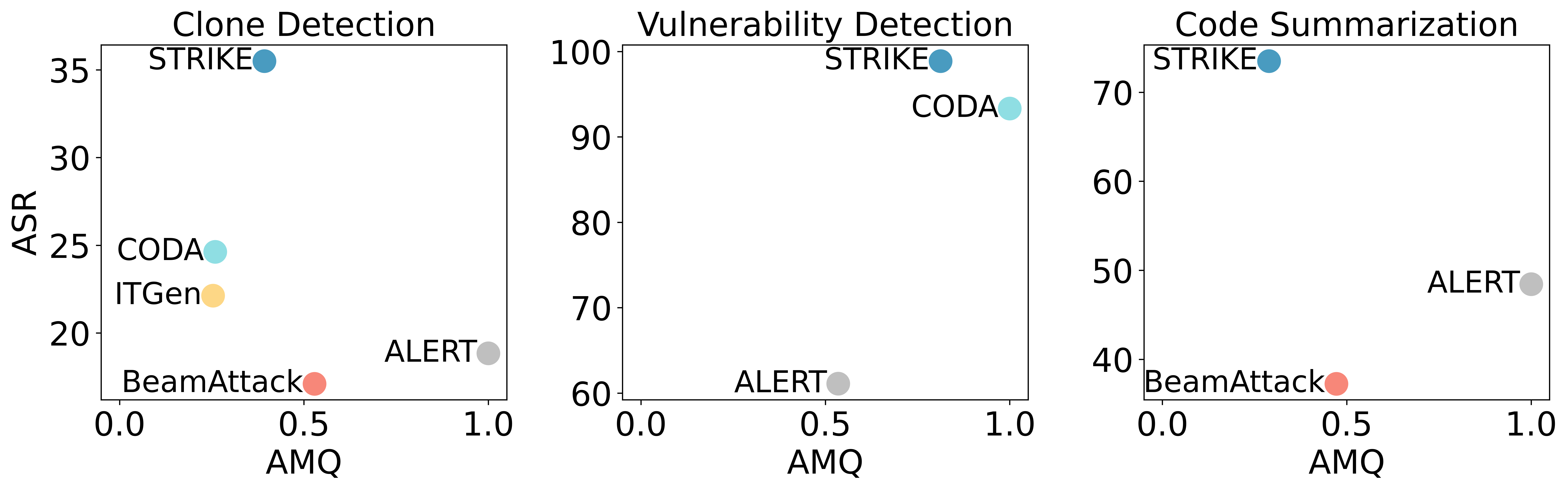}
    \end{minipage}
    \vspace{-2mm} 
    \begin{minipage}{\linewidth}
        \centering
        \includegraphics[width=\linewidth]{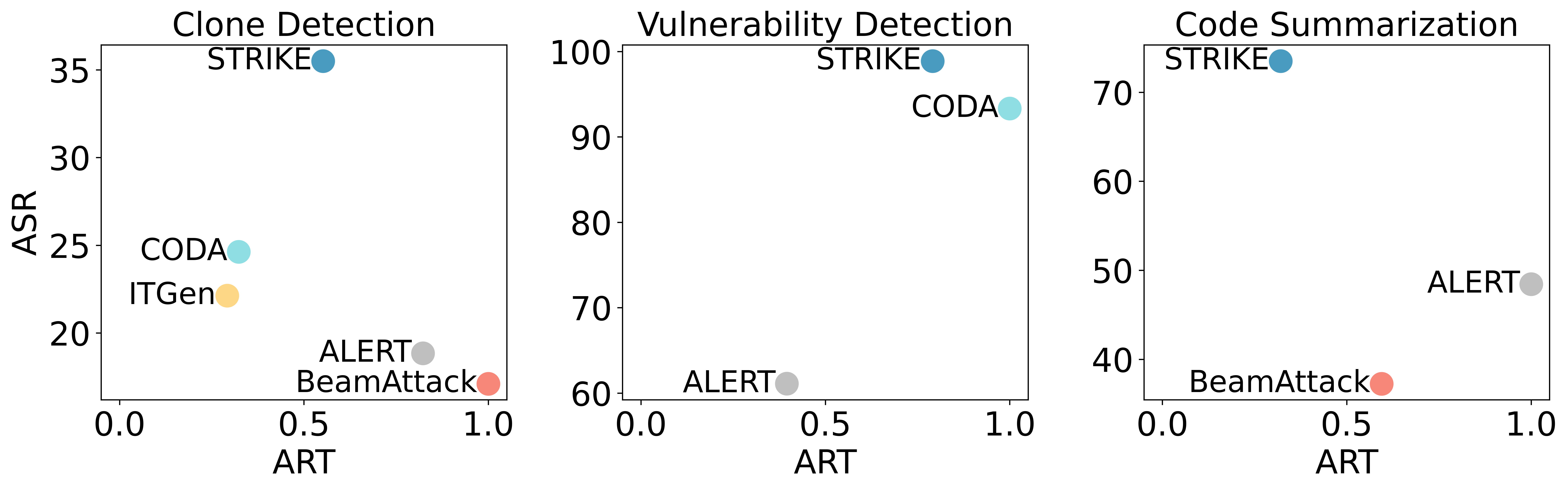}
    \end{minipage}
    \caption{ Effectiveness~($\uparrow$) -- Efficiency~($\leftarrow$) Analysis.}
    \label{fig:pareto_three}
    \vspace{-2mm}
\end{figure}

\begin{table*}[t!]
\centering
\footnotesize
\caption{ACS of Different Attack Methods on Three Tasks.}
\vspace{-2mm}
\label{tab:acs}
\setlength{\tabcolsep}{2pt}
\renewcommand{\arraystretch}{0.5}

\begin{tabular}{l c c c c c| c c c| c c c}
\toprule
\multirow{3}{*}{\textbf{Model}} 
& \multicolumn{5}{c}{\textbf{Clone Detection}} 
& \multicolumn{3}{c}{\textbf{Vulnerability Detection}}
& \multicolumn{3}{c}{\textbf{Code Summarization}} \\

\cmidrule(lr){2-6} \cmidrule(lr){7-9} \cmidrule(lr){10-12}

& \tool & CODA & ITGen & BeamAttack & ALERT
& \tool & CODA & ALERT
& \tool & BeamAttack & ALERT \\

\midrule

\textbf{CodeBERT} 
& 96.74\% & 93.54\% & 96.38\% & \textbf{98.56\%} & 96.72\%
& 96.98\% & 93.47\% & \textbf{97.69\%}
& 96.78\% & \textbf{98.19\%} & 97.47\% \\

\textbf{CodeGPT} 
& 96.73\% & 93.30\% & 94.73\% & \textbf{98.55\%} & 96.44\%
& 97.65\% & 93.71\% & \textbf{97.91\%}
& 97.07\% & \textbf{98.92\%} & 97.51\% \\

\textbf{CodeT5} 
& 96.92\% & 95.30\% & 95.50\% & \textbf{97.96\%} & 96.18\%
& \textbf{97.50\%} & 93.03\% & 97.41\%
& 97.63\% & \textbf{98.07\%} & 97.00\% \\

\midrule

\textbf{Average}
& 96.80\% & 94.05\% & 95.54\% & \textbf{98.36\%} & 96.45\%
& 97.38\% & 93.40\% & \textbf{97.67\%}
& 97.16\% & \textbf{98.39\%} & 97.33\% \\

\bottomrule
\end{tabular}
\vspace{-3mm}
\end{table*}

\subsubsection{Efficiency}
We measure attack efficiency using AMQ and ART. AMQ counts target-model queries, while ART measures the end-to-end attack time, including candidate generation and subsequent adversarial search. Lower values indicate lower query or runtime overhead.
Fig.~\ref{fig:AMQ_ART} reports efficiency in terms of ART (top) and normalized AMQ (bottom). For ART, ITGen and CODA show the lowest runtime on clone detection, followed by \tool, while BeamAttack and ALERT incur higher overhead due to larger search exploration. On vulnerability detection, \tool\ remains competitive, CODA is the slowest, and ALERT is the most time-efficient. On code summarization, \tool\ achieves the lowest runtime and ALERT the highest. Overall runtimes remain within a practical range, with \tool\ introducing only moderate overhead. 
For normalized AMQ, ITGen and CODA require fewer queries on clone detection, followed by \tool, whereas BeamAttack and ALERT consume noticeably more queries. On vulnerability detection, \tool\ remains competitive, CODA incurs the highest query overhead, and ALERT becomes the most efficient. On code summarization, \tool\ achieves the lowest AMQ, with BeamAttack second and ALERT still exhibiting the largest overhead. These results indicate that \tool\ balances attack effectiveness with target-model query and measured search-time overhead across the evaluated tasks.
Prior attack methods primarily focus on predefined identifier renaming or injecting structural blocks from other samples, resulting in homogeneous adversarial samples or limited perturbation diversity and contextual naturalness.

\noindent\textbf{\textit{LLM Request Cost.}}
AMQ captures target-model query cost, which is the primary query resource in the score- and output-based black-box setting.
\tool\ additionally incurs auxiliary LLM inference for structural candidate generation.
In the main experiments, \tool\ requires an average of 5.4, 3.7, and 4.2 auxiliary LLM requests per attacked input for clone detection, vulnerability detection, and code summarization, respectively.
OpenChat-3.5 is executed locally on the RTX 6000 Ada GPU, and each request generates all $n{=}10$ candidate rewrites for one identified code block.
ART includes this local LLM inference time, together with candidate parsing, filtering, target-model inference, and adversarial search.
The identifier-substitution stage does not invoke the LLM.
On code summarization, the lower target-model query count and early termination offset the auxiliary candidate-generation time, resulting in the lowest measured ART among the evaluated methods.

\subsubsection{Effectiveness-Efficiency}
Fig.~\ref{fig:pareto_three} compares the average ASR and AMQ/ART of three models across three tasks. 
Points closer to the upper-left indicate higher attack effectiveness with lower query and runtime overheads, where \tool\  appears nearest to this region.
Although ITGen and CODA require fewer queries and runtimes, they achieve lower ASR on clone detection, while \tool\ attains the highest success rate with only moderate overhead, yielding a better effectiveness–efficiency trade-off.
BeamAttack and ALERT remain in the lower-right region across all tasks, indicating both weaker attack strength and higher overhead.
Overall, \tool\ provides a favorable empirical trade-off between attack success, target-model queries, and measured search time in the evaluated settings.



\subsection{RQ2: Behavioral Preservation, Similarity, and Naturalness}

We evaluate the generated adversarial samples using complementary forms of evidence. Strategy-specific static checks are applied across all three tasks. Execution-based behavioral validation is additionally conducted on the executable Juliet vulnerability-detection subset. Representation similarity and human evaluation assess whether the generated variants remain close to the original code and contextually natural. 




\begin{table*}[t!]
\centering
\scriptsize
\caption{Robustness Analysis under Different Adversarial Attack Methods.}
\vspace{-2mm}
\label{tab:robustness}
\setlength{\tabcolsep}{1.5pt}
\renewcommand{\arraystretch}{1}
\begin{tabular}{l l|cccccc|cccc|cccc}
\toprule
\multirow{4}{*}{\textbf{Method}} & \multirow{4}{*}{\textbf{Model}}
& \multicolumn{6}{c|}{\textbf{Clone Detection}} 
& \multicolumn{4}{c|}{\textbf{Vulnerability Detection}}
& \multicolumn{4}{c}{\textbf{Code Summarization}} \\
\cmidrule(lr){3-8} \cmidrule(lr){9-12} \cmidrule(lr){13-16}

& 
& Clean & \tool & CODA & ITGen & Beam & ALERT
& Clean & \tool & CODA & ALERT
& Clean & \tool & Beam & ALERT \\

\midrule

\multirow{4}{*}{\tool-FT}
& CodeBERT 
& \textbf{96.97\%} & \textbf{90.57\%} & \textbf{87.15\%} & 93.84\% & 84.86\% & 88.22\%
& \textbf{64.35\%} & \textbf{64.94\%} & \textbf{56.91\%} & \textbf{64.08\%}
& \textbf{19.80} & \textbf{11.61} & \textbf{14.93} & \textbf{14.14} \\

& CodeGPT
& \textbf{97.62\%} & \textbf{89.78\%} & \textbf{93.58\%} & \textbf{95.91\%} & \textbf{91.65\%} & \textbf{91.42\%}
& \textbf{64.28\%} & \textbf{72.08\%} & \textbf{62.84\%} & \textbf{68.93\%}
& \textbf{20.06} & \textbf{9.85} & \textbf{12.11} & {12.35} \\

& CodeT5
& \textbf{97.25\%} & \textbf{92.24\%} & \textbf{90.85\%} & 94.20\% & 89.13\% & \textbf{92.57\%}
& \textbf{63.87\%} & \textbf{68.83\%} & \textbf{60.68\%} & \textbf{65.70\%}
& \textbf{20.47} & \textbf{12.34} & \textbf{14.82} & \textbf{14.44} \\

& Average
& \textbf{97.28\%} & \textbf{90.86\%} & \textbf{90.53\%} & \textbf{94.65\%} & \textbf{88.55\%} & \textbf{90.74\%}
& \textbf{64.17\%} & \textbf{68.62\%} & \textbf{60.14\%} & \textbf{66.24\%}
& \textbf{20.11} & \textbf{11.27} & \textbf{13.95} & \textbf{13.64} \\

\midrule

\multirow{4}{*}{CODA-FT}
& CodeBERT 
& 96.10\% & 86.95\% & 86.49\% & 84.04\% & 73.17\% & 79.21\%
& 64.16\% & 58.09\% & 54.71\% & 55.34\%
& -- & -- & -- & -- \\

& CodeGPT  
& 96.20\% & 85.08\% & 82.54\% & 85.90\% & 70.33\% & 72.68\%
& 63.34\% & 62.17\% & 58.90\% & 67.73\%
& -- & -- & -- & -- \\

& CodeT5   
& 96.60\% & 84.95\% & 89.79\% & 85.63\% & 83.71\% & 84.65\%
& 63.32\% & 45.82\% & 53.31\% & 48.12\%
& -- & -- & -- & -- \\

& Average
& 96.30\% & 85.66\% & 86.27\% & 85.19\% & 75.74\% & 78.85\%
& 63.61\% & 55.36\% & 55.64\% & 57.06\%
& -- & -- & -- & -- \\

\midrule

\multirow{4}{*}{ITGen-FT}
& CodeBERT 
& 96.87\% & 82.79\% & 78.10\% & 92.65\% & 92.16\% & 92.85\%
& -- & -- & -- & -- 
& -- & -- & -- & -- \\

& CodeGPT  
& 97.00\% & 71.06\% & 72.74\% & 69.70\% & 70.16\% & 78.29\%
& -- & -- & -- & -- 
& -- & -- & -- & -- \\

& CodeT5   
& 96.92\% & 79.22\% & 70.79\% & 93.41\% & \textbf{95.74\%} & 90.11\%
& -- & -- & -- & -- 
& -- & -- & -- & -- \\

& Average
& 96.93\% & 77.69\% & 73.88\% & 85.25\% & 86.02\% & 87.08\%
& -- & -- & -- & -- 
& -- & -- & -- & -- \\

\midrule

\multirow{4}{*}{Beam-FT}
& CodeBERT 
& 96.90\% & 67.25\% & 61.54\% & 94.36\% & \textbf{94.19\%} & 89.53\%
& -- & -- & -- & -- 
& 18.91 & 11.23 & 13.24 & 14.03 \\

& CodeGPT  
& 96.72\% & 75.19\% & 75.27\% & 81.02\% & 71.65\% & 76.70\%
& -- & -- & -- & -- 
& 19.70 & 9.48 & 11.73 & \textbf{13.57} \\

& CodeT5   
& 97.05\% & 82.74\% & 88.71\% & \textbf{97.08\%} & 90.30\% & 90.19\%
& -- & -- & -- & -- 
& 20.45 & 10.73 & 13.91 & 13.08 \\

& Average
& 96.89\% & 75.06\% & 75.17\% & 90.82\% & 85.38\% & 85.47\%
& -- & -- & -- & -- 
& 19.69 & 10.48 & 12.96 & 13.56 \\

\midrule

\multirow{4}{*}{ALERT-FT}
& CodeBERT 
& 96.12\% & 63.36\% & 64.60\% & \textbf{97.65\%} & 93.69\% & \textbf{95.41\%}
& 64.20\% & 47.41\% & 43.09\% & 54.44\%
& 18.97 & 10.93 & 14.72 & 12.21 \\

& CodeGPT  
& 97.20\% & 71.03\% & 65.88\% & 83.45\% & 68.34\% & 80.84\%
& 63.26\% & 58.50\% & 52.98\% & 58.08\%
& 19.58 & 9.46 & 12.03 & 12.70 \\

& CodeT5   
& 97.10\% & 77.06\% & 69.14\% & 89.94\% & 91.19\% & 90.70\%
& 63.17\% & 46.62\% & 40.81\% & 58.85\%
& 19.57 & 10.92 & 14.23 & 13.87 \\

& Average
& 96.81\% & 70.48\% & 66.54\% & 90.35\% & 84.41\% & 88.98\%
& 63.54\% & 50.84\% & 45.63\% & 57.12\%
& 19.37 & 10.44 & 13.66 & 12.93 \\

\bottomrule
\end{tabular}
\vspace{-2mm}
\end{table*}

\begin{figure}[t!]
    \centering
    \includegraphics[width=0.97\linewidth]{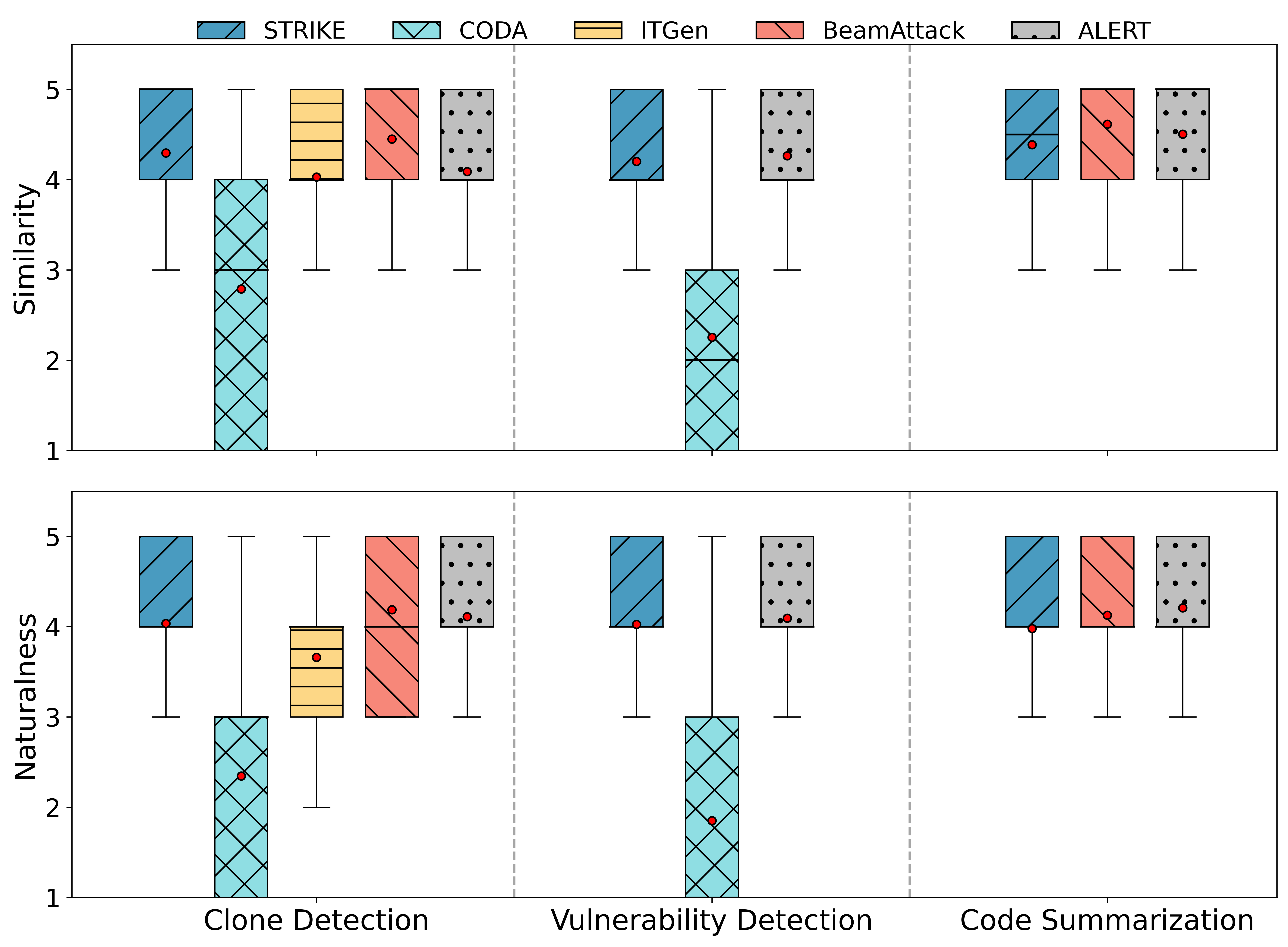}
    \vspace{-2mm}
    \caption{Subjective Analysis Results.}
    \label{fig:human_eval}
    \vspace{-4mm}
\end{figure}

\subsubsection{Objective Analysis}
We use ACS to measure similarity in embedding space, as it captures how closely an adversarial code segment aligns with the original representation. Higher ACS values indicate stronger similarity in embedding space. As shown in Table~\ref{tab:acs}, differences emerge across perturbation strategies. On clone detection, \tool\ achieves high ACS, with an average of 96.80\%, which is slightly lower than the highest ACS method, BeamAttack, and higher than others. Although CODA also performs structural and identifier perturbations, its reliance on transferring code blocks from opposite-label samples occasionally introduces stylistic drift or unintended structural deviation, leading to the lowest ACS. In contrast, \tool\ maintains high embedding similarity by leveraging LLM-based structural perturbations and similarity-guided identifier substitutions. On vulnerability detection and code summarization tasks, \tool\ likewise achieves high ACS, comparable to other methods. These results suggest that while identifier-oriented attacks naturally achieve high embedding similarity, \tool\ maintains comparable similarity while introducing more effective adversarial perturbations.

\subsubsection{Subjective Analysis}
Following prior works~\cite{tian2023coda,yang2022alert}, we assess whether samples remain similar and natural to the original code from a human perspective.

\noindent\textbf{\textit{Data Preparation.}}
For each task, we randomly select 10 instances and one successful adversarial sample per method, forming pairs of original and adversarial code. 
In total, we evaluate five methods on clone detection and three methods on each of the other two tasks (11 method–task combinations); with 10 samples and 3 models per combination, yielding $11 \times 10 \times 3 = 330$ pairs.

\noindent\textbf{\textit{Process.}}
All participants hold at least a Master’s degree in Computer Science and have over five years of programming experience. 
The study was approved by the institutional ethics committee. 
Adversarial methods are anonymized and randomly shuffled to ensure objectivity. 
Each pair is rated on two aspects: (1) \textit{contextual naturalness}~\cite{hindle2012naturalness,rabin2021generalizability}, evaluating whether perturbations remain developer-like and contextually consistent; and (2) \textit{similarity}~\cite{tian2023coda,yefet2020adversarial}, assessing whether the adversarial sample remains similar to the original code in overall structure and human-viewed plausibility.  
Ratings use a 5-point Likert scale (1 = strongly unsatisfied, 5 = strongly satisfied).

\noindent\textbf{\textit{Results.}}
We collected ratings from 12 non-author evaluators, each assessing all 330 pairs across three tasks and three models. Inter-rater reliability measured by Krippendorff’s $\alpha$ for ordinal data~\cite{krippendorff2011computing} is 0.679, indicating acceptable agreement. As shown in Fig.~\ref{fig:human_eval}, \tool\ achieves high scores in contextual naturalness, with distributions concentrated near the upper range~(red dots indicate the mean rating). For similarity, \tool\ performs comparably to identifier-based methods such as ALERT and BeamAttack. 
In contrast, CODA receives lower scores on both metrics, likely because inserting code blocks from opposite-label samples can disrupt structural flow and indentation. Overall, the results indicate that the structure-aware perturbations generated by \tool\ produce natural and human-perceived similar adversarial code, consistent with the ACS results.

\subsubsection{Behavioral Preservation}

\noindent\textbf{\textit{Static Checks.}}
We apply a post-generation static audit to every final candidate before counting it as a successful attack, using Tree-sitter ASTs and the transformation records produced during attack generation. For Dead Code Insertion, we check that all original statements remain unchanged and that newly inserted statements do not assign to pre-existing identifiers, invoke functions, or contain control-transfer statements. For Control Structure Replacement, we conservatively compare the original and transformed control structures with respect to their conditions, body statements, side-effecting expressions, loop initialization and update expressions, and explicit control-transfer statements. Cases whose behavioral preservation cannot be established conservatively are marked as unverified. For Statement Reordering, we require the statement multiset to remain unchanged and preserve the relative order of statements with apparent def-use dependencies. For Identifier Substitution, we check that the recorded renaming is applied consistently to all occurrences within the corresponding lexical scope without introducing name collisions or capture. Each sample is classified as passed, failed, or unverified, and we report the corresponding distributions for the four perturbation strategies. These checks provide conservative conformance evidence rather than a proof of semantic equivalence. The pass/fail/unverified distributions are 98.72\%/0.61\%/0.67\% for Dead Code Insertion, 92.64\%/1.83\%/5.53\% for Control Structure Replacement, 97.91\%/0.96\%/1.13\% for Statement Reordering, and 99.58\%/0.19\%/0.23\% for Identifier Substitution.

\noindent\textbf{\textit{Execution-Based Validation.}}
For vulnerability detection, we map each evaluated function back to its complete Juliet source file using the source-file path and function name retained during dataset preprocessing. We replace only the target function with its perturbed version, while keeping the remaining source files, test harness, inputs, and build configuration unchanged. The original and perturbed versions are compiled and executed using the same compiler, instrumentation, environment, and timeout. We execute each original test case three times and retain only cases with a deterministic and observable runtime verdict. For vulnerable cases, behavioral preservation requires that the perturbed version retain the same sanitizer or runtime-failure category observed for the original version. For benign cases, both versions must terminate normally without sanitizer findings, abnormal termination, or timeout, and must produce the same normalized output. Cases that cannot be mapped, compiled, executed, or assigned a reliable runtime verdict are treated as unvalidated and excluded from the BPR, raw-ASR, and confirmed-ASR denominators. We report validation coverage as the proportion of eligible cases included in this validated subset. We compute raw ASR and confirmed ASR on the same validated subset, with confirmed ASR counting only successful attacks that preserve the original runtime verdict.

\begin{table*}[t]
\centering
\footnotesize
\caption{Execution-Based Behavioral Validation on Juliet.}
\label{tab:behavior_validation}
\setlength{\tabcolsep}{3pt}
\renewcommand{\arraystretch}{0.85}
\begin{tabular}{lrrrrrr}
\toprule
\textbf{Model} &
\textbf{Eligible} &
\textbf{Validated} &
\textbf{Coverage} &
\textbf{BPR} &
\textbf{Raw ASR} &
\textbf{Confirmed ASR} \\
\midrule
CodeBERT & 1,742 & 1,538 & 88.29\% & 98.61\% & 98.82\% & 97.57\% \\
CodeGPT  & 1,782 & 1,571 & 88.16\% & 98.78\% & 98.71\% & 97.62\% \\
CodeT5   & 1,738 & 1,526 & 87.80\% & 98.39\% & 97.83\% & 96.35\% \\
\bottomrule
\end{tabular}
\end{table*}

As shown in Table~\ref{tab:behavior_validation}, 87.80\%--88.29\% of the eligible Juliet cases could be mapped, compiled, executed, and assigned a deterministic runtime verdict. Within this validated subset, 98.39\%--98.78\% of the generated variants preserved the original runtime verdict. When attack success was additionally conditioned on runtime-verdict preservation, the confirmed ASR remained 96.35\%--97.62\%, compared with a raw ASR of 97.83\%--98.82\% on the same validated subset.

\begin{table*}[t]
\centering
\footnotesize
\caption{Ablation Study of \tool\ on LLMs.}
\vspace{-2mm}
\label{tab:ablation_llm}
\setlength{\tabcolsep}{1.2pt}
\renewcommand{\arraystretch}{0.7}

\begin{tabular}{l|l|cccc|cccc|cccc}
\toprule
\multirow{2}{*}{\textbf{Model}} 
& \multirow{2}{*}{\textbf{Setting}} 
& \multicolumn{4}{c|}{\textbf{Clone Detection}} 
& \multicolumn{4}{c|}{\textbf{Vulnerability Detection}} 
& \multicolumn{4}{c}{\textbf{Code Summarization}} \\

\cmidrule(lr){3-6} \cmidrule(lr){7-10} \cmidrule(lr){11-14}

&
& ASR$\uparrow$ & AMQ$\downarrow$ & ART$\downarrow$ & ACS$\uparrow$
& ASR$\uparrow$ & AMQ$\downarrow$ & ART$\downarrow$ & ACS$\uparrow$
& ASR$\uparrow$ & AMQ$\downarrow$ & ART$\downarrow$ & ACS$\uparrow$ \\

\midrule

\multirow{2}{*}{\textbf{CodeBERT}}
& \textbf{\tool} 
& \textbf{41.98\%} & \textbf{905.95} & \textbf{0.35} & 96.74\%
& \textbf{98.88\%} & 1505.39 & \textbf{0.33} & 96.89\%
& \textbf{89.98\%} & 155.33 & \textbf{0.14} & 96.78\% \\

& \textbf{\tool-GPT5} 
& 41.18\% & 917.71 & 0.47 & \textbf{96.76\%}
& 98.68\% & \textbf{1464.72} & 0.43 & \textbf{97.12\%}
& 88.95\% & \textbf{152.35} & 0.22 & \textbf{97.19\%} \\

\midrule

\multirow{2}{*}{\textbf{CodeGPT}}
& \textbf{\tool} 
& 32.45\% & 932.27 & \textbf{0.41} & \textbf{96.73\%}
& \textbf{98.82\%} & 1409.20 & 0.29 & \textbf{97.65\%}
& 66.31\% & \textbf{241.17} & \textbf{0.50} & 97.07\% \\

& \textbf{\tool-GPT5} 
& \textbf{34.69\%} & \textbf{815.86} & 0.48 & 96.15\%
& 91.54\% & \textbf{332.40} & \textbf{0.12} & 97.52\%
& \textbf{70.97\%} & 272.84 & 0.55 & \textbf{97.40\%} \\

\midrule

\multirow{2}{*}{\textbf{CodeT5}}
& \textbf{\tool} 
& 32.07\% & 915.66 & 0.84 & 96.92\%
& \textbf{97.91\%} & 697.64 & 0.41 & 97.50\%
& 64.25\% & \textbf{160.98} & \textbf{0.38} & 97.63\% \\

& \textbf{\tool-GPT5} 
& \textbf{39.00\%} & \textbf{799.20} & \textbf{0.71} & \textbf{96.95\%}
& 96.95\% & \textbf{338.25} & \textbf{0.25} & \textbf{97.87\%}
& \textbf{73.12\%} & 246.99 & 0.52 & \textbf{97.83\%} \\

\bottomrule
\end{tabular}

\vspace{-2mm}
\end{table*}

\subsection{RQ3: Model Robustness}

We examine cross-attack robustness after fine-tuning with samples generated by different attacks.
Following prior adversarial fine-tuning protocols~\cite{tian2023coda,huang2025itgen}, we partition each task's test set into two disjoint subsets, $S_1$ and $S_2$.
Each attack is applied to $S_1$ to construct attack-specific augmentation data, pairing each input with either a successful adversarial sample or, if no successful sample is found, the variant that produces the largest reduction in the task-specific attack score.
The models are then fine-tuned on the original training data combined with the corresponding augmentation set.
We follow prior settings~\cite{du2023beam,tian2023coda,huang2025itgen}, using learning rates of $5{\times}10^{-5}$, $2{\times}10^{-5}$, and $5{\times}10^{-5}$ and training for 2, 5, and 10 epochs for clone detection, vulnerability detection, and code summarization, respectively.
$S_2$ is not used during fine-tuning or model selection.
For evaluation, each attack is applied to $S_2$, and its successful variants are retained to form fixed attack-induced evaluation sets.
This protocol measures cross-attack performance on these fixed subsets rather than robustness over the full original test distribution.
We report performance on clean data (to verify no degradation) and on each adversarial set, measuring cross-attack robustness, where higher values indicate better preservation of model performance under adversarial attacks.

\noindent\textbf{\textit{Results.}}
As shown in Table~\ref{tab:robustness}, \tool-fine-tuned models preserve clean performance across all tasks, indicating no noticeable degradation on clean data.
On Vulnerability Detection, \tool\ achieves the strongest robustness with clear margins over all baselines across attacks, demonstrating effective resistance to diverse perturbations. 
For Clone Detection, while some baselines perform best under their own attack distributions, \tool\ maintains high performance across all attacks, indicating stronger cross-attack generalization. 
For Code Summarization, \tool\ provides consistent but smaller gains, suggesting that improving robustness for generative tasks remains more challenging. 
Overall, \tool-based fine-tuning improves cross-attack robustness on fixed attack-induced evaluation subsets, while preserving clean performance.

\subsection{RQ4: Ablation Study}

\begin{table}[t]
\centering
\footnotesize
\caption{Ablation Study of \tool\ on Modules.}
\vspace{-2mm}
\label{tab:ablation_asr}
\setlength{\tabcolsep}{3pt}
\renewcommand{\arraystretch}{0.7}

\begin{tabular}{lcccccc}
\toprule
\textbf{Model} & \textbf{\tool} & \textbf{w/o {DCI}} & \textbf{w/o {CSR}} & \textbf{w/o {SR}} & \textbf{w/o SLP} & \textbf{w/o ILS} \\
\midrule
\textbf{CodeBERT} & \textbf{76.95\%} & 67.04\% & 69.76\% & 71.36\% & 63.04\% & 28.89\% \\
\textbf{CodeGPT}  & \textbf{65.86\%} & 61.38\% & 58.63\% & 60.55\% & 53.23\% & 29.31\% \\
\textbf{CodeT5}   & \textbf{64.74\%} & 58.15\% & 55.76\% & 57.31\% & 48.38\% & 37.52\% \\
\bottomrule
\end{tabular}
\vspace{-4mm}
\end{table}

\begin{figure*}[t!]
  \centering
  \includegraphics[width=0.97\linewidth]{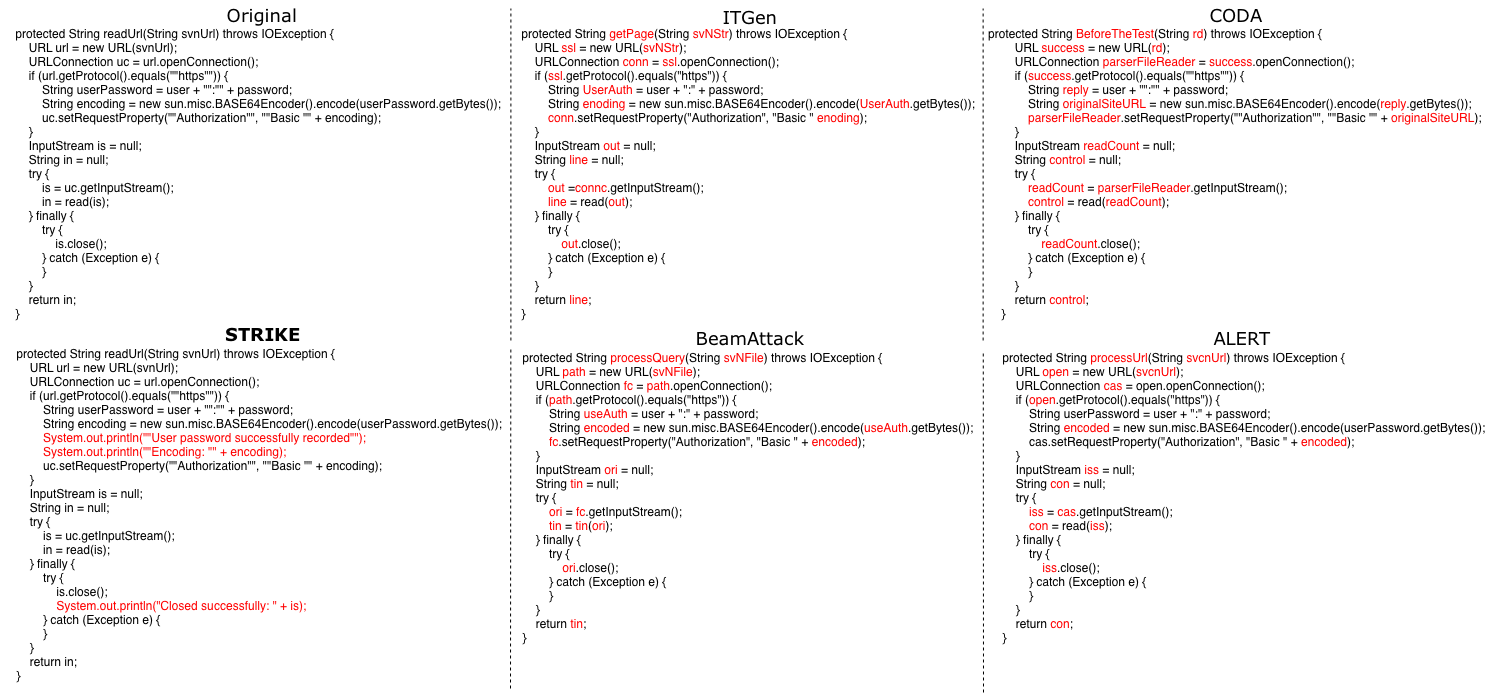}
  \vspace{-4mm}
  \caption{Example~(\#519 in BigCloneBench/testset) of Adversarial Code: Baselines vs. \tool.}
  \label{fig:example-6}
  \vspace{-2mm}
\end{figure*}

\subsubsection{LLMs}

We analyze the influence of different LLMs during the perturbations in \tool, and design two variants:

\begin{itemize}[leftmargin=1.0em]
    \item \textbf{\tool}: uses the open-source LLM, OpenChat-3.5~\cite{wang2023openchat}, by default, enabling easy reproducibility.
    \item \textbf{\tool-GPT5}: replaces the LLM used in structure-level perturbations with GPT-5-nano~\cite{openai_gpt5nano}, to assess whether more advanced LLMs yield stronger adversarial performance.
\end{itemize}
This ablation examines how the choice of LLM backend affects the performance of \tool. As shown in Table~\ref{tab:ablation_llm}, the impact of the backend is model- and task-dependent. On CodeBERT, \tool\ and \tool-GPT5 produce comparable results. On CodeGPT and CodeT5, \tool-GPT5 sometimes achieves higher ASR and lower AMQ/ART. And \tool-GPT5 generally yields slightly higher ACS, suggesting that a stronger LLM can produce perturbations with higher embedding similarity to the original code. \tool\ remains effective under both backends, while the choice of backend can influence the attack effectiveness and efficiency in some settings. These results show that \tool\ is compatible with both LLM backends, although the choice of backend affects attack effectiveness and efficiency in some settings.

\begin{figure*}[t!]
    \centering
    \includegraphics[width=0.97\linewidth]{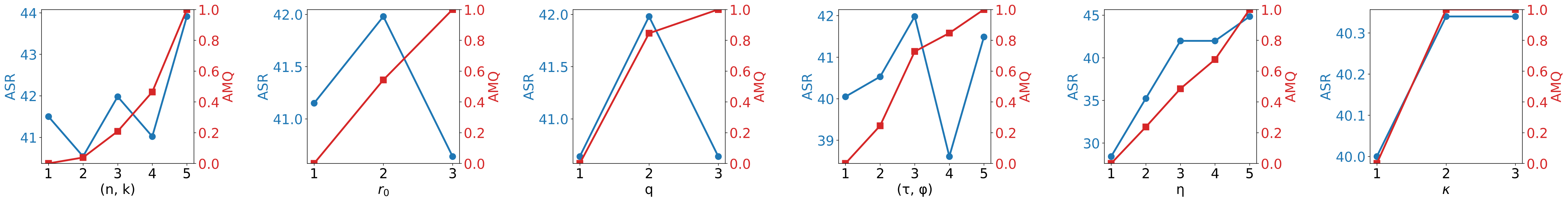}
    \vspace{-3mm}
    \caption{Impact of Hyperparameters.}
    \label{fig:hyper}
    \vspace{-4mm}
\end{figure*}

\subsubsection{Modules}

We examine the contribution of each main module in \tool, and construct five ablated variants:

\begin{itemize}[leftmargin=1.0em]
    \item \textbf{w/o {DCI}}: remove dead code insertion.
    
    \item \textbf{w/o {CSR}}:  remove control structure replacement.
    
    \item \textbf{w/o {SR}}: remove statement reordering.

    \item \textbf{w/o SLP}: remove all structure-level perturbations.
    
    \item \textbf{w/o ILS}: remove identifier-level substitutions.
\end{itemize}
Following previous work~\cite{tian2023coda}, these ablations isolate the effect of each structural component and the identifier substitution stage, allowing us to quantify the incremental contribution of each component.
Table~\ref{tab:ablation_asr} shows the average ASR across tasks on CodeBERT, CodeGPT, and CodeT5.
\tool\ consistently outperforms all ablated variants, demonstrating the effectiveness of its modular design.
Among structural stages, removing DCI, CSR, or SR individually leads to moderate ASR drops, indicating that each contributes complementary diversity to the perturbation space.
When all structure-level perturbations are removed, the ASR clearly decreases, confirming their collective role in expanding perturbation diversity.
In contrast, eliminating identifier-level substitution causes a larger performance drop, indicating that DCMs are highly sensitive to identifier-level signals. However, identifier substitutions alone quickly saturate because they generate highly similar lexical variants within a narrow search space. Structural perturbations are therefore essential: by introducing diverse program variants and modifying structures, they expand the adversarial search space while preserving high similarity to the original code, enabling subsequent identifier substitutions to trigger more effective attacks.

\section{Discussion}

\subsection{Perturbation of Adversarial Code}
Fig.~\ref{fig:example-6} shows an example: \tool\ caused misclassification after only 152 queries during dead code insertion, while all baselines failed even after {674–2783} queries. Baselines produced largely homogeneous edits, mostly identifier renames, whereas \tool’s LLM-based structural perturbation succeeds with higher ACS (98.18\%) than ALERT~(98.14\%), BeamAttack~(97.83\%), ITGen~(96.36\%), and CODA~(97.28\%). The successful perturbations consist of inserting a small number of developer-like lines within the selected block while remaining close to the original code in representation similarity and contextual plausibility. This helps explain why \tool\ achieves stronger attack effectiveness in this example. Compared with prior methods, such perturbations also provide different adversarial samples for adversarial fine-tuning, contributing to improved cross-attack robustness while preserving clean-task performance.

\subsection{Impact of Hyperparameters in \tool}

We evaluate the sensitivity of \tool's hyperparameters on clone detection using CodeBERT.
For the structural stage, $(n,k)$ control the number of LLM-generated candidates and the retained subset, $r_0$ controls the active search width, and $q$ controls the patience threshold. We test 5 settings: $(n,k)\!\in\!\{(4,2),(6,3),(10,5),(20,10),(40,20)\}$, and 3 settings: $r_0\!\in\!\{2,3,4\}$ and $q\!\in\!\{1,2,3\}$.
For identifier substitution, $(\tau,\phi)$ determine the number of sampled external code and the least similar code candidates, and $\eta$ controls the most similar identifier candidates, and $\kappa$ control threshold.
We test 5 settings: $(\tau,\phi)\!\in\!\{(64,8),(128,16),(256,32),(512,64),(1024,128)\}$ and $\eta\!\in\!\{10,20,30,40,50\}$, and three settings: $\kappa\!\in\!\{1,2,3\}$. As these hyperparameters affect ACS only marginally (within ±0.2\%) and ART trends closely follow AMQ, we focus our analysis on ASR and AMQ. 
As shown in Fig.~\ref{fig:hyper}, each x-axis index corresponds to one predefined setting (5 settings for $(n,k)$, $(\tau,\phi)$, and $\eta$, and 3 settings for $r_0$, $q$, and $\kappa$). The results show that ASR stabilizes around the default settings. Further increasing parameter values yields only marginal ASR gains, while AMQ rises sharply, by 2–4 × in some cases, making larger settings computationally inefficient.

\subsection{Additional Recent DCMs}

We further evaluate \tool\ on two more recent DCMs, StarCoderBase-1B~\cite{li2023starcoder, starcoderbase1b} and Qwen2.5-Coder-1.5B~\cite{hui2024qwen2, qwen25coder15b}, for vulnerability detection. Following the main protocol, we compare \tool\ with CODA, the strongest applicable baseline in the main vulnerability-detection experiments, using the same originally correctly classified samples.

\begin{table}[t]
\centering
\footnotesize
\caption{Results on Additional Recent DCMs.}
\vspace{-1mm}
\label{tab:recent_models}
\setlength{\tabcolsep}{3.8pt}
\renewcommand{\arraystretch}{0.8}
\begin{tabular}{ll|ccccc}
\toprule
\textbf{Model} &
\textbf{Method} &
\textbf{Acc.} &
\textbf{ASR$\uparrow$} &
\textbf{AMQ$\downarrow$} &
\textbf{ART$\downarrow$} &
\textbf{ACS$\uparrow$} \\
\midrule
\multirow{2}{*}{StarCoderBase-1B}
& \tool & 61.63\% & 99.56\% & 638.91 & 0.19 & 97.72\% \\
& CODA  & 61.63\% & 95.43\% & 1,462.37 & 0.46 & 93.68\% \\
\midrule
\multirow{2}{*}{Qwen2.5-Coder-1.5B}
& \tool & 61.86\% & 99.96\% & 806.02 & 0.21 & 97.26\% \\
& CODA  & 61.86\% & 96.18\% & 1,548.62 & 0.52 & 93.21\% \\
\bottomrule
\end{tabular}
\vspace{-4mm}
\end{table}

\tool\ achieves a higher ASR than CODA on both models while maintaining high code similarity, providing additional evidence on two recent code-model architectures in the vulnerability-detection setting.

\subsection{Threats to Validity}

\noindent \textbf{Internal Validity.}
Variability in experimental results may arise from different hyperparameter configurations. To mitigate this, we conducted sensitivity analyses on hyperparameters in \tool. Another potential threat stems from implementation errors. To minimize this risk, we performed thorough code reviews and released the complete implementation and baselines for reproducibility. {Since most benchmark datasets consist of isolated code snippets rather than complete executable programs, execution-based validation is not universally applicable. We apply strategy-specific static checks across all three tasks and conduct execution-based differential validation only on the executable Juliet vulnerability-detection subset. Approximately 88\% of the eligible Juliet cases could be assigned a deterministic runtime verdict; the remaining cases could not be mapped, compiled, executed, or reliably classified and may introduce selection bias. For clone detection and code summarization, representation similarity and human evaluation provide complementary evidence but do not establish behavioral or semantic equivalence. In addition, BLEU-4=0 is used as a proxy for severe summarization degradation for consistency with prior work, but this lexical criterion may not fully capture semantic changes in generated summaries. Finally, our benchmark setting assumes access to ground-truth labels or reference summaries and should therefore be interpreted as offline robustness testing rather than a fully label-free attack against a deployed service.}

\noindent \textbf{External Validity.}
{Generalizability may be affected by the choice of tasks, datasets, and models. We mitigate this by evaluating \tool\ across multiple tasks, datasets, and models. \tool\ is not tied to fixed corpora and extends to new settings, with results validated on both open- and closed-source LLM backends. Additional results on Python datasets are included in our artifact.}

\section{Related Work}

Adversarial attacks on DCMs fall into two categories: black-box and white-box.

\noindent\textbf{Black-box attacks.}
These methods query model outputs to guide identifier- or structure-level edits. 
ALERT~\cite{yang2022alert} uses a masked language model and genetic search for identifier substitution, but is limited to renaming and can be query-intensive. 
BeamAttack~\cite{du2023beam} employs beam search guided by context and statement importance, improving efficiency but remaining sensitive to beam size and favoring shallow local edits. 
CODA~\cite{tian2023coda} combines structural and identifier transformations using opposite-label samples, but depends heavily on label information, cannot generalize to generative tasks, and often produces less natural or structurally inconsistent code. 
ITGen~\cite{huang2025itgen} formulates attack generation as a discrete black-box optimization problem using Bayesian optimization with ARD kernels and DPP sampling to improve query efficiency, but still focuses on identifier-level perturbations. 
Other methods~\cite{springer2020strata,jha2023codeattack} rely on heuristic search, yet are constrained by predefined transformations or task-specific signals, limiting similarity and robustness.

\noindent\textbf{White-box attacks.}
These techniques rely on gradients or internal representations to guide perturbation search.
DAMP~\cite{yefet2020adversarial} identifies influential variable substitutions using gradients, while CARROT~\cite{zhang2022carrot} prioritizes mutations via gradient saliency; code-obfuscation approaches~\cite{srikant2021adversarial,schloegel2022loki} generate samples through program transformations but often sacrifice generalizability and efficiency.

\noindent\textbf{Summary.}
Existing adversarial methods differ in search strategy and model access but share key limitations. White-box methods require gradient access, limiting generalizability. Black-box methods typically focus on identifier substitution, resulting in a narrow attack space and homogeneous adversarial code. CODA broadens the perturbation space through edits transferred from opposite-label reference samples, but its candidate space remains bounded by the reference pool and tied to classification labels. In contrast, \tool\ constructs structural candidates from the local context of the attacked program and complements them with dynamically generated identifier candidates, enabling the same framework to support both classification and generation tasks.

\section{Conclusion}

We presented \tool, an input-conditioned black-box adversarial robustness-testing framework that constructs and searches complementary structural and identifier perturbation spaces for each attacked program. Across three representative code intelligence tasks, \tool\ achieves higher attack success rates than the applicable baselines with competitive target-model query overhead. Strategy-specific static checks are applied across all tasks, while execution-based validation on the executable Juliet subset provides task-bounded evidence of behavioral preservation for vulnerability detection. Representation similarity and human evaluation further indicate that the generated variants remain close to the original code and contextually plausible, although these analyses do not establish semantic equivalence for clone detection or code summarization. Adversarial fine-tuning with \tool-generated samples also improves cross-attack performance on fixed perturbed evaluation sets while preserving clean performance. Overall, the results demonstrate that constructing the perturbation space from the attacked input, rather than relying only on identifier-level edits or opposite-label reference transfers, enables effective robustness testing across both classification and generation tasks.

\bibliographystyle{ACM-Reference-Format}
\bibliography{main}

\end{document}